\documentclass[aps,prl,reprint,floatfix,superscriptaddress]{revtex4-1}
\usepackage[utf8x]{inputenc}
\usepackage[english]{babel}

\usepackage{amsmath}
\usepackage{graphicx}
\usepackage{physics}
\usepackage{bm} 
\usepackage{dcolumn} 
\usepackage[colorlinks=true, citecolor=blue, linkcolor=blue]{hyperref}%

\renewcommand{\i}{{\rm i}}
\newcommand{\e}{\mathrm{e}}
\renewcommand{\d}{\mathrm{d}}

\draft 

\begin{document}

\title{Exciton energy oscillations induced by quantum beats} 

\author{A.~V.~Trifonov}
\email[correspondence address: ]{a.trifonov@spbu.ru}
\affiliation{Spin Optics Laboratory, St. Petersburg State University, 1 Ul’anovskaya str., Peterhof, St. Petersburg 198504, Russia}
\author{A. S. Kurdyubov}
\affiliation{Spin Optics Laboratory, St. Petersburg State University, 1 Ul’anovskaya str., Peterhof, St. Petersburg 198504, Russia}
\author{I. Ya. Gerlovin}
\affiliation{Spin Optics Laboratory, St. Petersburg State University, 1 Ul’anovskaya str., Peterhof, St. Petersburg 198504, Russia}
\author{D. S. Smirnov}
\affiliation{Ioffe Institute, St. Petersburg, 194021, Russian Federation}
\author{K. V. Kavokin}
\affiliation{Spin Optics Laboratory, St. Petersburg State University, 1 Ul’anovskaya str., Peterhof, St. Petersburg 198504, Russia}
\author{I. A. Yugova}
\affiliation{V.A. Fock Institute of Physics, St. Petersburg State University, 1 Ul’anovskaya str., Peterhof, St. Petersburg 198504, Russia}
\affiliation{Spin Optics Laboratory, St. Petersburg State University, 1 Ul’anovskaya str., Peterhof, St. Petersburg 198504, Russia}
\author{M. Aßmann}
\affiliation{Experimentelle Physik 2, Technische Universität Dortmund, Dortmund, D-44221, Germany}
\author{A. V. Kavokin}
\affiliation{Institute of Natural Sciences, Westlake University, No.18, Shilongshan Road, Cloud Town, Xihu District, Hangzhou, China
}
\affiliation{Spin Optics Laboratory, St. Petersburg State University, 1 Ul’anovskaya str., Peterhof, St. Petersburg 198504, Russia}
\date{\today}

\begin{abstract}
In this paper we experimentally demonstrate an oscillating energy shift of quantum-confined exciton levels in a semiconductor quantum well after excitation into a superposition of two quantum confined exciton states of different parity. Oscillations are observed at frequencies corresponding to the quantum beats between these states. We show that observed effect is a manifestation of the exciton density oscillations in the real space similar to the dynamics of a Hertzian dipole.
The effect is caused by the exciton-exciton exchange interaction and appears only if the excitons are in a superposition quantum state. Thus, we have found clear evidence for the incoherent exchange interaction in the coherent process of quantum beats. This effect may be harnessed for quantum technologies requiring the quantum coherence of states. 


\end{abstract}


\maketitle 

Excitons in high-quality semiconductor nanostructures are considered as one of the most promising systems for the implementation for quantum computation~\cite{Bennet_Nat.00,Chemla_Nat01}. Their high radiative decay rate (up to $10^{11}$~s$^{-1}$~\cite{Andreani_ssc91, DeveaudPRL_91}) makes it possible to operate such systems at excellent rates  and renders light-matter interaction highly efficient. Exciton systems are also characterised by strong non-linearities and long-lived, so-called Raman coherence which are an essential building block for many quantum computation architectures. It is well known that a major contribution to these strong non-linear properties is given by the exciton-exciton interaction~\cite{Schindler-PRB2008,Ciuti-PRB1998}. For example, it is the exciton-exciton scattering process that provides functioning of a polariton laser ~\cite{Savvidis-PRL2000, Deng-Science2002, Kasprzak-Nat2006}. Another striking manifestation of this interaction is the energy shift of exciton (polariton) resonances with increasing exciton density under  intense optical excitation~\cite{Peyghambarian_PRL84,Wachter_Physica02,Sie_NanoLet17}. Here, a long-term exciton spectral shift occurs due to the interaction of exciton states~\cite{VMAxt} with other excitations generated by light: free carriers~\cite{Schlaad_PRB91, Brown_PRB01} and excitons~\cite{Schmitt-Rink_PRB85, Wachter_Physica02}. 

Experiments show~\cite{Blackwood-PRB1994,Amo-PRB2010,Trifonov-PRL2019} that, in semiconductor quantum wells, the amplitude of the exciton energy shift strongly depends on the mutual orientation of the spins of the interacting excitons. According to Ref.~\cite{Ciuti-PRB1998}, the primary spin-dependent mechanism causing an exciton energy shift is the exchange interaction. In this case, the main contribution to the interaction energy arises due to the exchange of electrons and holes in the interacting excitons. The corresponding energy shift, $\delta\varepsilon$, equals to $\alpha N_{X}$, where $\alpha$ is the constant of the exchange interaction (exchange integral) and $N_{X}$ is the concentration of excitons in the system. Note, that the exchange integral does not depend on time, so up to date, the time dependence of the energy shift in  time-resolved experiments was only observed as an exponential decay of $\delta\varepsilon$ associated with the decay of $N_{X}$.

A prominent manifestation of quantum Raman coherence is the appearance of quantum beats caused by the interference of states of a quantum system. This effect may be observed under simultaneous excitation of several energy levels to a coherent superposition state. Quantum beats were experimentally observed as periodic oscillations of the intensity in time resolved luminescence~\cite{AleksandrovOS,DoddPPS1964}, or as oscillations of the amplitude in four-wave mixing and in  pump-probe experiments~\cite{MitsunagaPRA87,GobelPRL90,Schmitt-RinkPRB92, WundkePRB96,JoschkoPRL97,PalPRB03, QBpaper,Shahbook}. It is important to note that in exciton systems this effect is not related to the interaction between the excitons. However, simultaneous appearance of the exciton-exciton interaction and quantum beats has not been observed so far.

Here we report on the first observation of the incoherent exciton-exciton interaction manifesting in coherent quantum beats. In our experiments, we have observed the oscillations of exciton resonance energies in time induced by a short optical pulses. We demonstrate that the energy oscillations appears due to the exchange interaction of quantum confined excitons created in a superposition state.

\begin{figure}
 \begin{minipage}[h]{\linewidth}
 \includegraphics[width=\columnwidth]{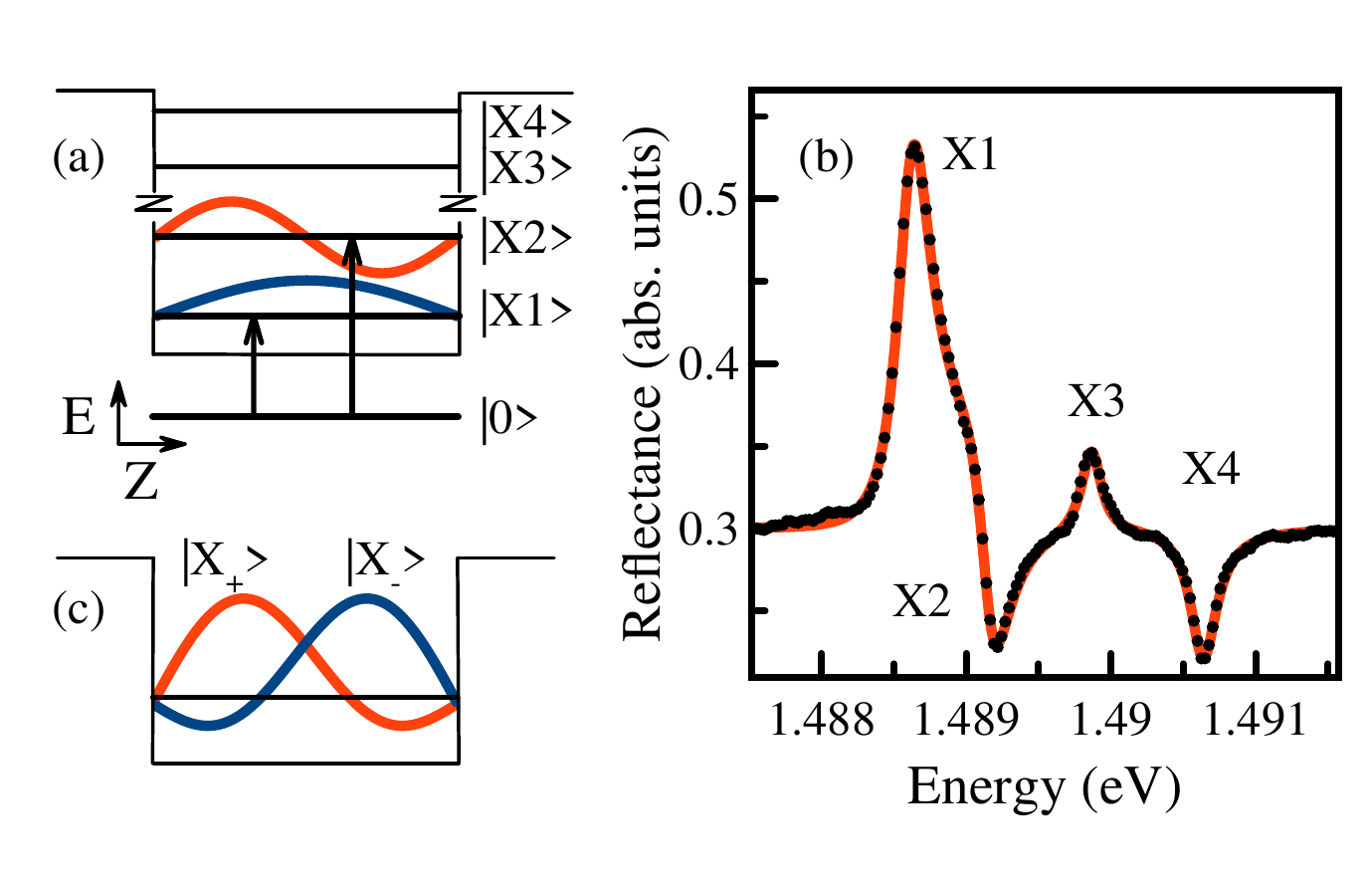}%
 \end{minipage}
 \caption{
 (a) Schematic of the potential profile of a wide quantum well showing the quantum-confined excitonic levels and corresponding envelopes of the exciton wave functions (coloured curves). (b) The experimental reflectance spectrum of the sample in the spectral range of quantum-confined exciton resonances (black dots) and the fit by Eqs.~\eqref{IvchenkoEq} and~\eqref{Reflectance} (red curve). Spectral features denoted as $X1$, ... $X4$, are associated with optical transitions from the ground state $\ket{0}$ to the corresponding quantum-confined exciton states. (c) Schematic of quantum well potential and envelopes of the wavefunctions $\ket{X_+}$ and $\ket{X_-}$ as defined in Eq.~\eqref{eq:Xpm}.} 
 \label{Fig1}
 \end{figure}

We experimentally study the nonlinear exciton dynamics in a sample with a relatively wide InGaAs quantum well. The sample was grown by molecular beam epitaxy  on a GaAs substrate. The sample contains a 90 nm thick In$_x$Ga$_{1-x}$As/GaAs quantum well (QW) with Indium concentration $x \approx 2.5 \%$. The depth of the QW for excitons is of about $25$~meV.
Figure~\ref{Fig1}(a) represents the potential profile of the quantum well for excitons and the quantum-confined exciton levels $\ket{X1}$, ... $\ket{XN}$. 
These states manifest themselves in the reflection spectrum of the sample as it is shown in Figure~\ref{Fig1}(b)~\cite{TRifonovNonTriv}. This system is well adapted for studies of processes related to quantum coherence since by varying the spectrum of a short optical pulse one may excite the system into various superposition states~\cite{QBpaper}.

\begin{figure}
 	\begin{minipage}[h]{\linewidth}
 		\center{\includegraphics[width=0.95\columnwidth]{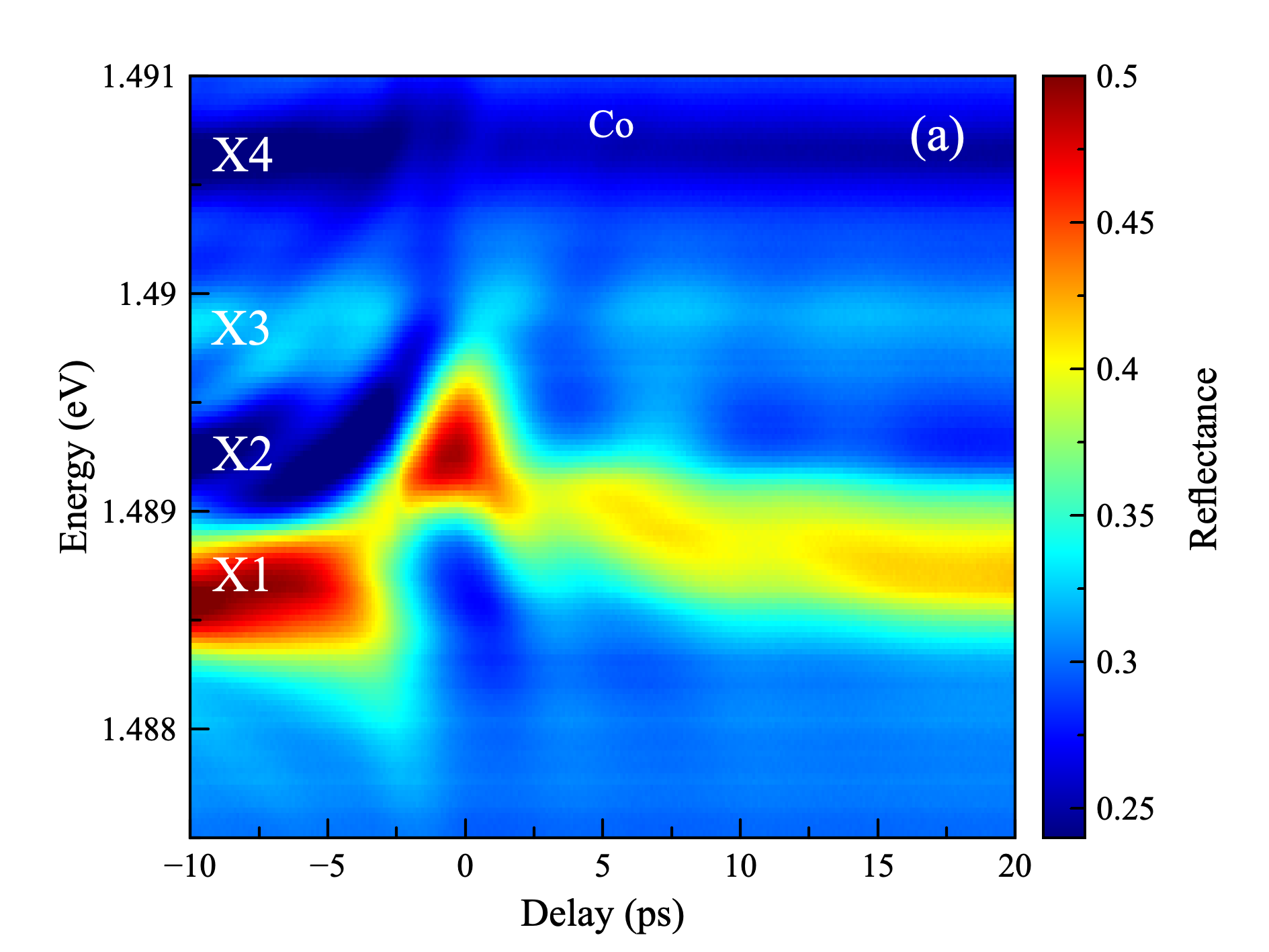}}
	\end{minipage}
  	\begin{minipage}[h]{\linewidth}
		 \center{\includegraphics[width=0.95\columnwidth]{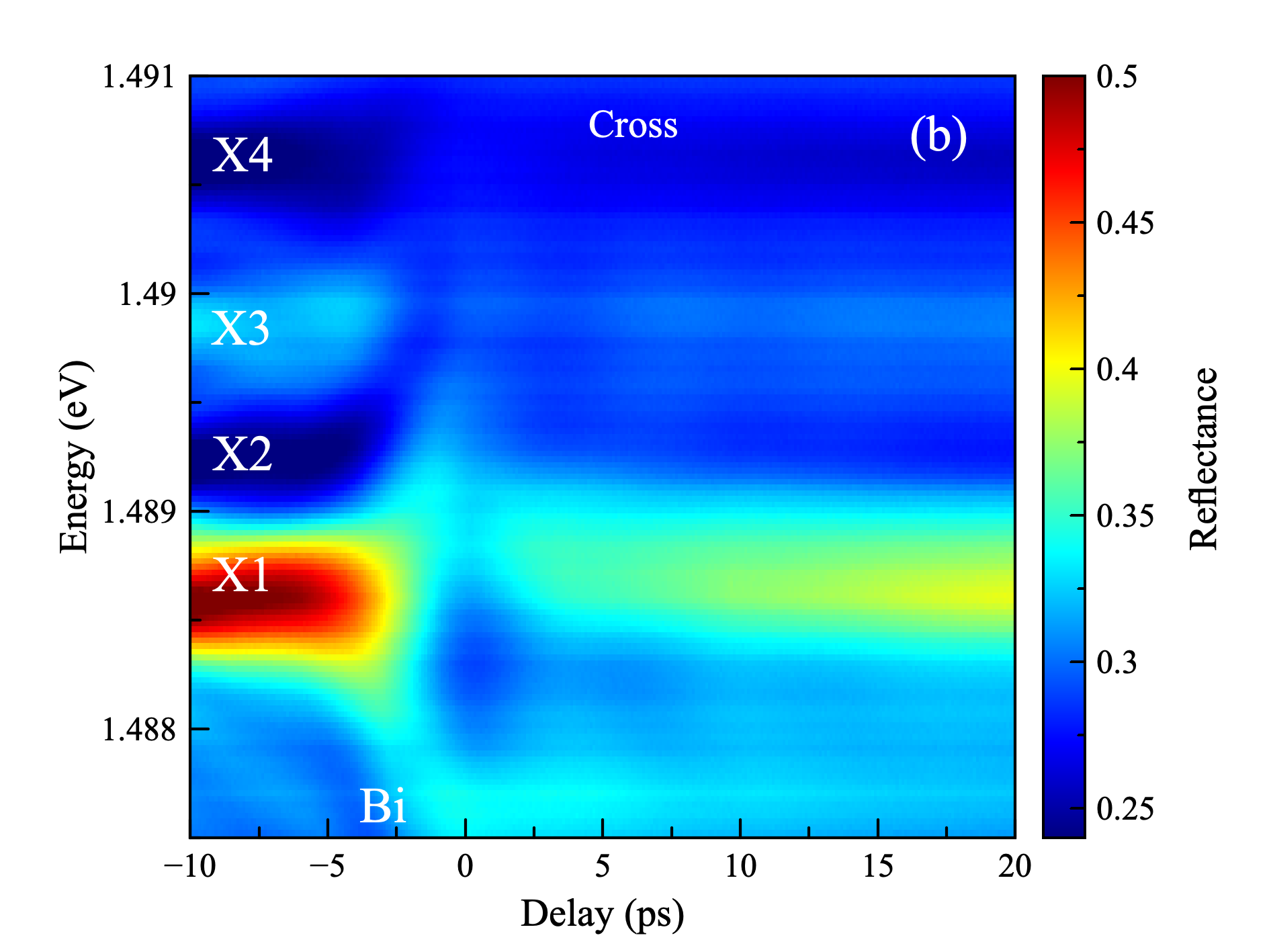}}
	\end{minipage}
 \caption{ Delay dependences of reflectance spectra in the circular (a) co - and (b) cross-polarization (with respect to the pump beam). Measurements were carried out at relatively high pump power ($12$~mW per $100$~$\mu$m spot). White inscriptions label the spectral features associated with quantum-confined exciton resonances.} 
 \label{Fig2}
 \end{figure}

The experiments were performed by spectrally and polarization-resolved pump-probe spectroscopy. In our experiments, we used  spectrally broad 100 fs probe pulses and spectrally narrow 2 ps pump pulses~\cite{Suppl,Trifonov-PRL2019}. The circularly polarised pump spectrally tuned to the exciton resonance X1 covered also the nearest exciton state X2, thereby exciting the system to a superposition state. The reflection spectra in co - and cross-circular polarizations for each delay between the pump and probe pulses were measured simultaneously. Figure~\ref{Fig2} shows the experimental data obtained at a relatively large pump power of 12 mW for a 100 $\mu$m spot .

Let us emphasise some of the most remarkable phenomena observed at positive delays between pump and probe pulses. The formation of the pump-probe signal at negative delays occurs due to a different mechanism~\cite{Lindberg-JOSA1988,Sokoloff- PRB1988,Lange-PRB2009}, its discussion is beyond the scope of this work. Experimental data obtained at co-polarized pump and probe beams (see  Figure~\ref{Fig2}(a)) reveal oscillations which manifest themselves not only in the amplitude but also in the energy  of the exciton resonances. Figure~\ref{Fig2}(b) shows that these oscillations are absent in cross-polarization. To obtain the parameters of the observed oscillations, we performed a detailed analysis of the reflection spectra. The analysis is based on the method presented in~\cite{IvchenkoBook}, which we generalised to the case of several nearest exciton states~\cite{TRifonovNonTriv,Khramtsov-PRB2019}. The amplitude reflection coefficient from the QW considering several exciton resonances reads:
\begin{equation}
r_{QW} = \sum\limits_{N=1}^{4} \frac{i(-1)^{N-1}\Gamma_{0N}e^{i\varphi_N}}%
{\tilde{\omega}_{0N} - \omega - i(\Gamma_{0N}+\Gamma_N)}.
\label{IvchenkoEq}
\end{equation}
Here $\tilde{\omega}_{0N}$ is the resonance frequency of the exciton resonance XN, $\Gamma_{0N}$ and $\Gamma_N$ are the radiative and non-radiative decay rates, respectively. The phase $\varphi_N$ is related to an asymmetry of the QW potential, caused by segregation of Indium during the growth process~\cite{Grigoryev-SM2016}. The reflectivity $R(\omega)$ of a structure with a cover layer of thickness $L_b$ and QW thickness $L_{QW}$  is given by~\cite{IvchenkoBook}:
\begin{equation}
R(\omega) = \left|\frac{r_{01}+r_{QW}e^{2i\phi}}{1+r_{01}r_{QW}e^{2i\phi}}\right|^2,
\label{Reflectance}
\end{equation}
where $r_{01}$ is the amplitude reflection coefficient of the sample surface, phase $\phi = K(L_b + L_{QW}/2)$, with $K$ being the wave vector of the photon in the heterostructure.

The reflectance spectra were fitted using  equations (\ref{IvchenkoEq}) and (\ref{Reflectance}). Despite a large number of parameters, all of them are uniquely determined with high accuracy. A set of reflection spectra measured at varying time delays between the pump and probe pulses was processed. For the processing, we used the results obtained at the relatively small pump power of $2$~mW per $100$~$\mu$m spot (6 times smaller than that used for the data presented in Figure~\ref{Fig2}, corresponding colourmaps shown in Suppl. Mat.~\cite{Suppl}). At higher pump powers, the shift of the exciton lines exceeded the distance between the X1 and X2 resonances, which did not allow us to separate these features spectrally.

\begin{figure}
 \begin{minipage}[h]{0.49\linewidth}
 \center{\includegraphics[width=\columnwidth]{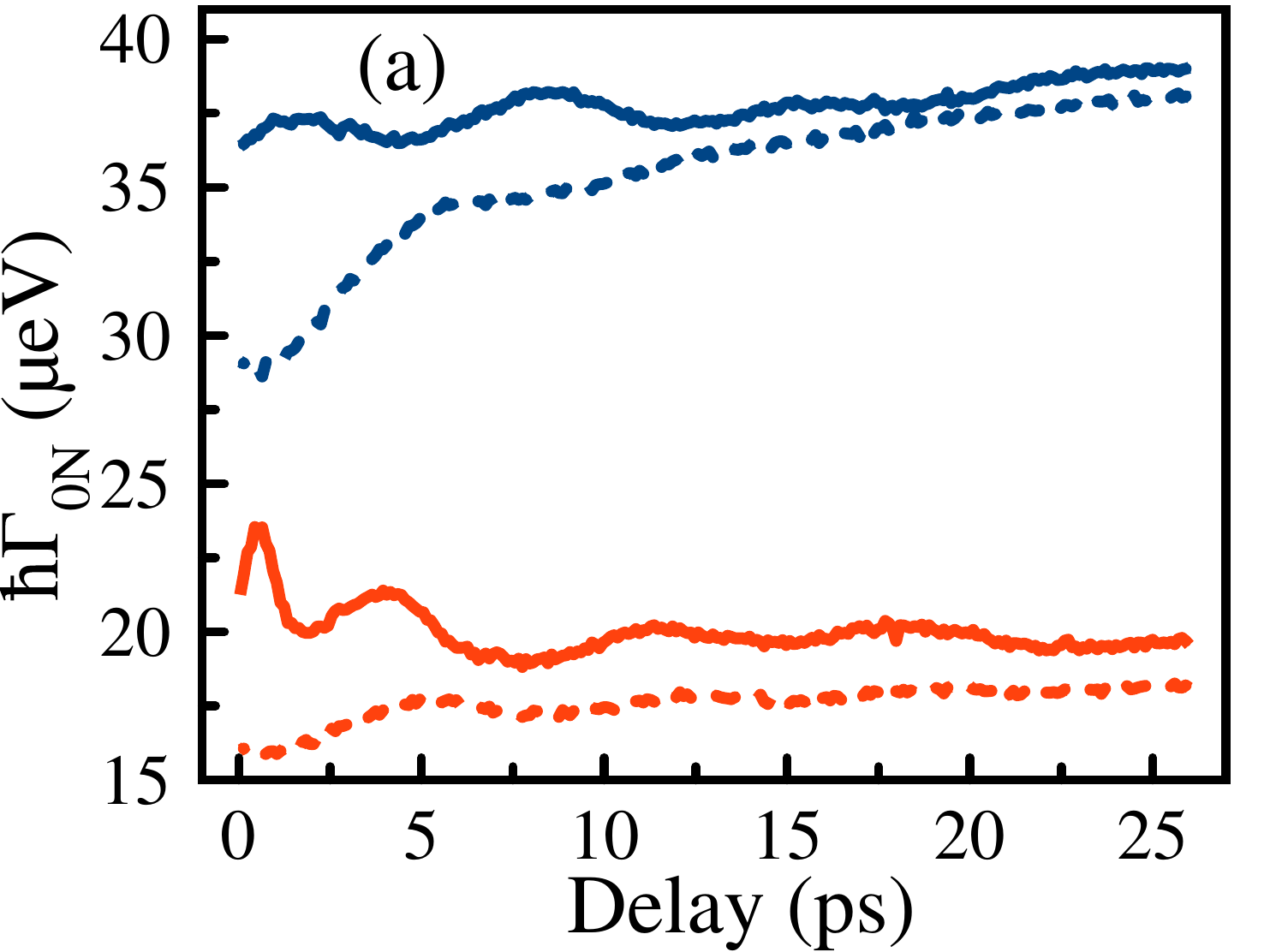}}
 \end{minipage}
 \hfill
  \begin{minipage}[h]{0.49\linewidth}
 \center{\includegraphics[width=\columnwidth]{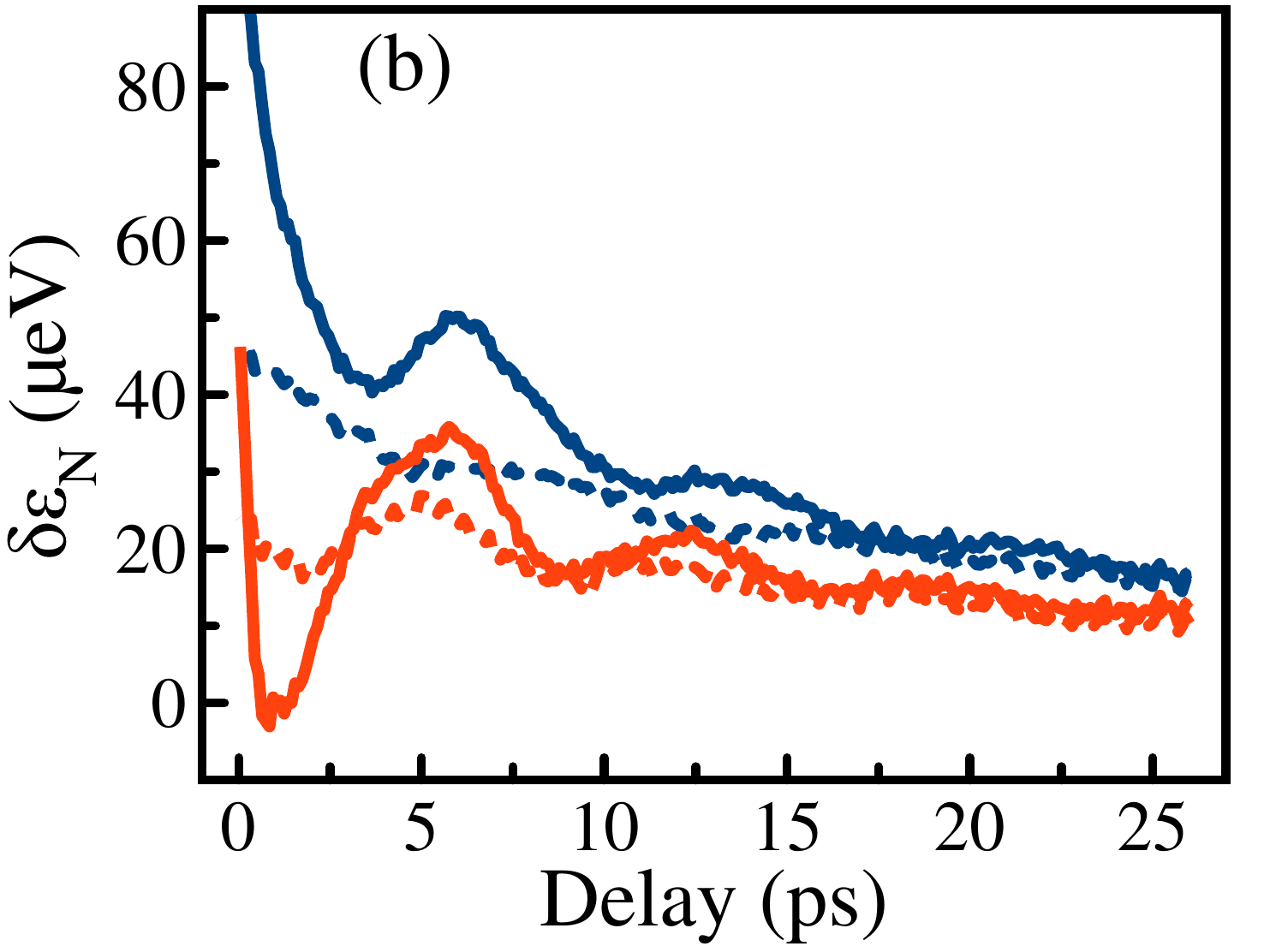}}
 \end{minipage}
 \vfill
   \begin{minipage}[h]{0.49\linewidth}
 \center{\includegraphics[width=\columnwidth]{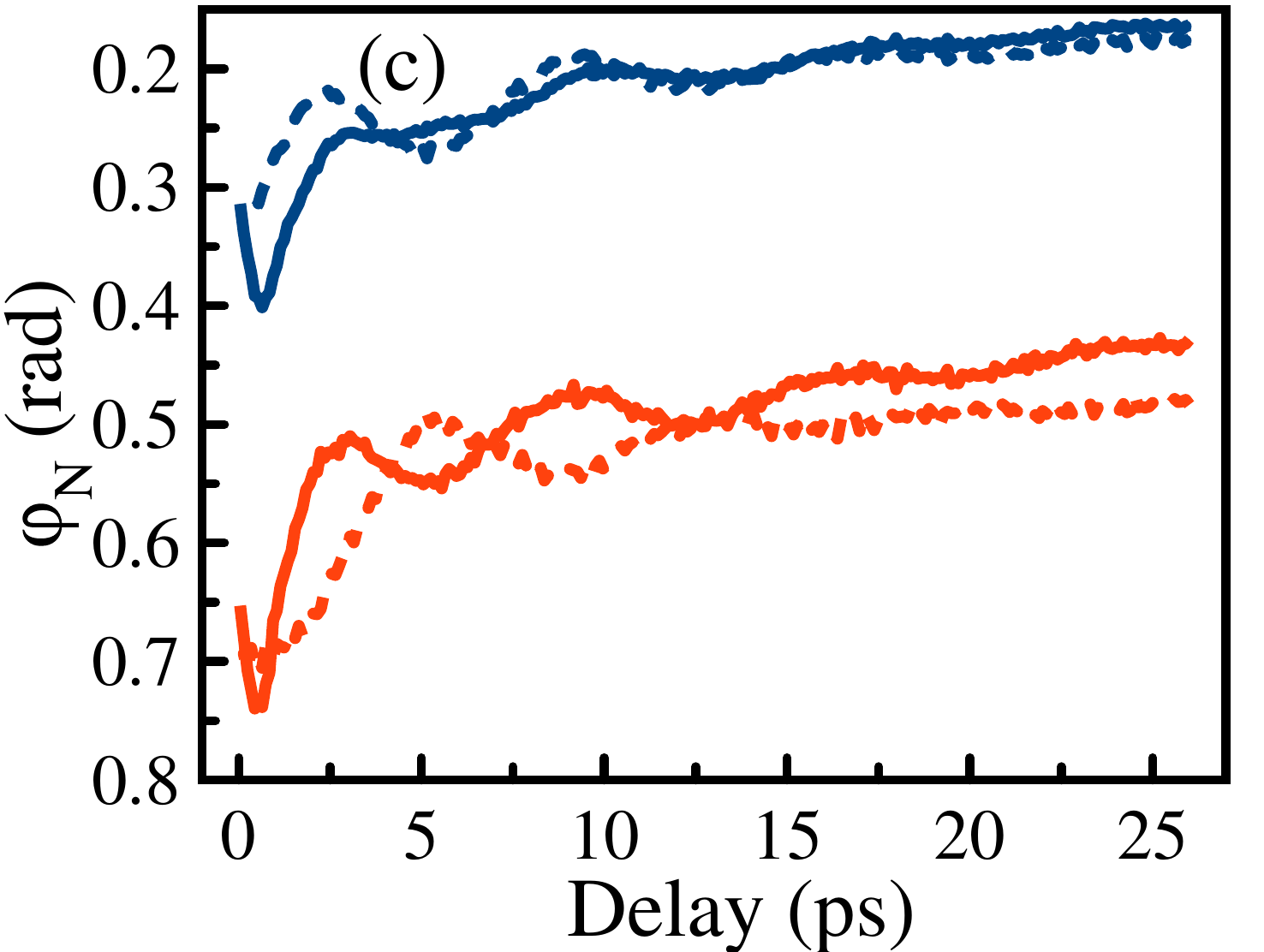}}
 \end{minipage}
 \hfill
   \begin{minipage}[h]{0.49\linewidth}
 \center{\includegraphics[width=\columnwidth]{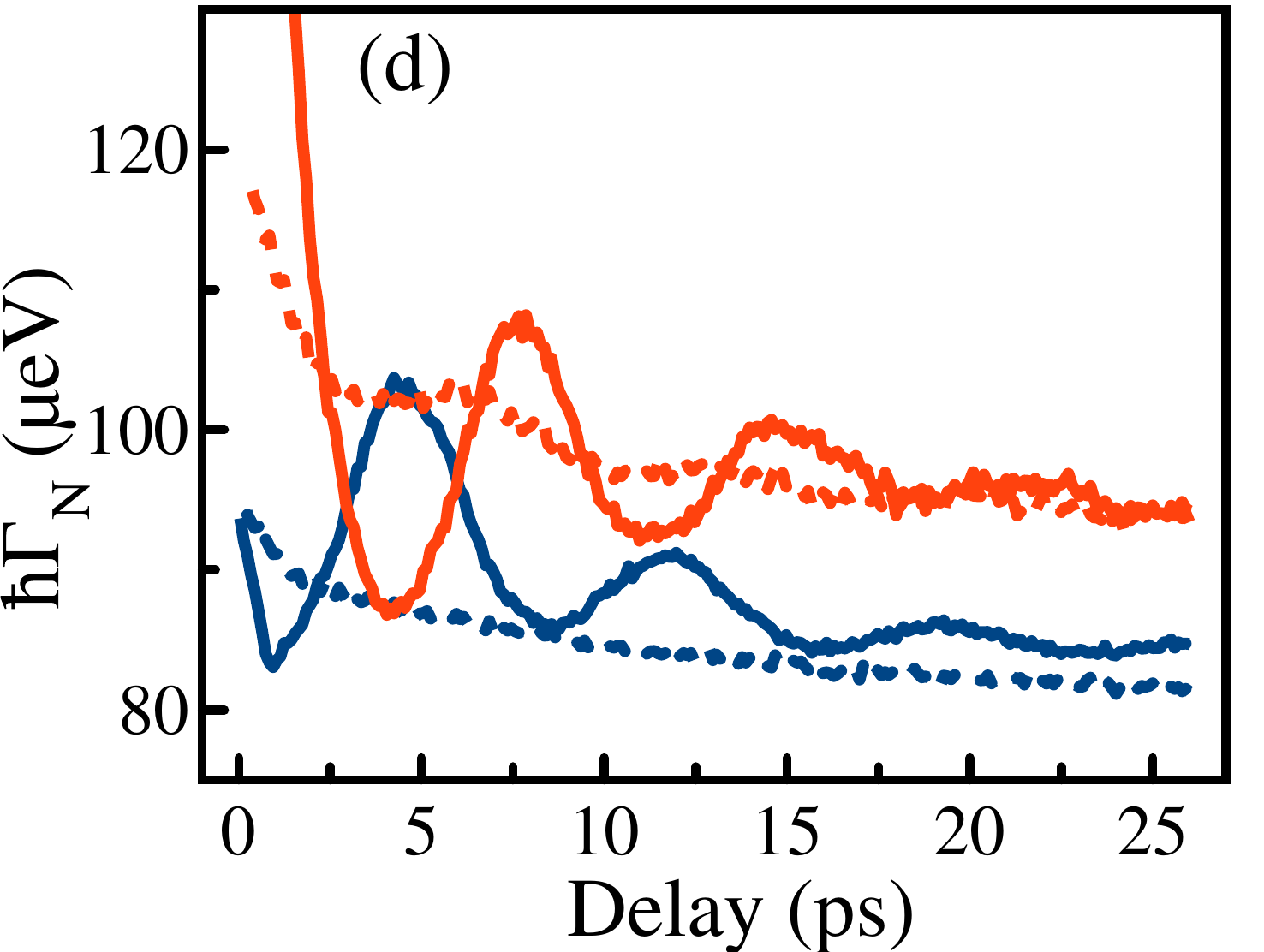}}
 \end{minipage}
 \caption{Delay dependence of the parameters of the exciton resonances X1 (blue) and X2 (red) ($N=1,2$) in the reflection spectra in co- and cross-polarizations (solid and dashed curves respectively). (a) the radiative broadening $\Gamma_{0N}$, (b) energy shift $\delta\varepsilon_N$, (c) the relative phase $\varphi_N$, (d) non-radiative broadening $\Gamma_{N}$.
 } 
 \label{FitParam}
 \end{figure}

Figure~\ref{FitParam} shows the delay dependences of the parameters for the X1 and X2 exciton resonances in the reflection spectra. The corresponding dependencies for the remaining excitonic resonances are presented in Suppl. Mat.~\cite{Suppl}.
These dependencies highlight the common origin of the observed phenomena. The delay oscillations of all parameters of the exciton resonance are clearly observed only for co-polarized pump and probe pulses. Importantly, the oscillations frequency corresponds to the energy difference between the exciton states X1 and X2, $\hbar (\tilde{\omega}_{02}-\tilde{\omega}_{01}) = 0.53$~meV.

Usually, the oscillations in the pump-probe signal, when the pump excites several exciton states, are explained in terms of oscillations of the total dipole moment of a three-level (multilevel) system, coherently excited to a superposition state. This is a well-known effect of quantum beats~\cite{AleksandrovOS,DoddPPS1964}, where the oscillations arise due to interference of electrical dipole moments of optical transitions oscillating at optical frequencies. It gives rise to oscillations of $\Gamma_0$ [see Figure~\ref{FitParam}(a)]. However, as our experiments show, the oscillations also manifest themselves in all other parameters of the exciton resonances. An oscillating dependence of the exciton resonance energies on the excitation to the superposition state has not been observed before. Thus, the obtained data provides a new insight into the interplay between the interaction in many-body ensembles and quantum beats phenomenon.

The oscillations observed in the delay dynamics of the exciton resonance frequency $\tilde{\omega}_{0N}$ are not connected with the interference of exciton dipole moments, but appear due to the exchange interaction between excitons excited to the superposition states. The crucial role of the exchange interaction is indicated by the pronounced dependence of the effect on the spin state of the excitations: oscillations of the exciton resonance parameters are clearly seen in the co-polarised probe and pump pulses and are absent for crossed polarizations. We believe that the small residual effect observed in cross-polarization can be traced back to biexciton formation. However, in this work we focus on the physical mechanism of formation of oscillations of exciton resonance energies observed in co-polarisation. 

The energy oscillations are caused by the quantum coherence between excitonic states X1 and X2, which can be described conveniently in the basis of the superposition states
  \begin{equation}
    \label{eq:Xpm}
    \ket{X_\pm} = \frac{\ket{X1}\pm\ket{X2}}{\sqrt{2}}.
  \end{equation}
For simplicity, let us assume equal dipole moments of $X_1$ and $X_2$ excitons and a spectrally broad pump pulse, while for the general case the results are qualitatively the same~\cite{Suppl}. In this limit at $t=0$ the pump pulse coherently excites $N_0$ excitons in the state $X_+$ only, and then the quantum beats between $X_+$ and $X_-$ states take place. The numbers of the excitons in these states are
\begin{equation}
  \label{eq:beats}
  N_\pm(t)=\frac{N_0}{2}\left[e^{-\gamma_0 t}+\cos\left(\Omega t\right)\e^{-\gamma_c t}\right],
\end{equation}
where $\Omega=\tilde\omega_{02}-\tilde\omega_{01}$ is the frequency of the beats, $\gamma_0$ is the exciton decay rate and $\gamma_c$ is the rate of decay of the Raman coherence between X1 and X2 states~\cite{Suppl}.

The probe pulse arrives at $t=\tau$ and creates a few $X_+$ excitons with a small in-plane wave vector, which interact with the pump excitons. We assume that interaction between the excitons created by the pump and probe pulses is proportional the the envelope of their wave functions. The envelope functions of $X_+$ and $X_-$ excitons are shown in Figure~\ref{Fig1}(c), where one can see, that they are shifted to the opposite sides of the QW. The repulsion energy between the excitons at the same side of the QW is $\omega_{ex}$, while the interaction between the excitons at the opposite sides of the QW can be neglected. As a result, the amplitudes of the probe excitons $X_\pm$ obey the Schr\"odinger equation
  \begin{equation}
    \label{eq:alpha}
    \frac{\d\alpha_\pm(t)}{\d t}=-\i\left[\omega_0+\omega_{ex}N_\pm(t)-\i\frac{\gamma_c}{2}\right]\alpha_\pm(t)+\i\frac{\Omega}{2}\alpha_\mp(t),
  \end{equation}
respectively, with the oscillating resonance frequencies $\omega_0+\omega_{ex}N_\pm(t)$. From the solution of these equations with the initial conditions $\alpha_+(\tau)=\alpha_0$ and $\alpha_-(\tau)=0$ one finds the amplitudes of the probe excitons and the probe excitons polarization $P(t)\propto\alpha_+(t)$. Ultimately, the Fourier transform of the polarization evolution provides the reflectivity spectrum of the QW.

\begin{figure}
 \begin{minipage}[h]{0.49\linewidth}
 \center{\includegraphics[width=\columnwidth]{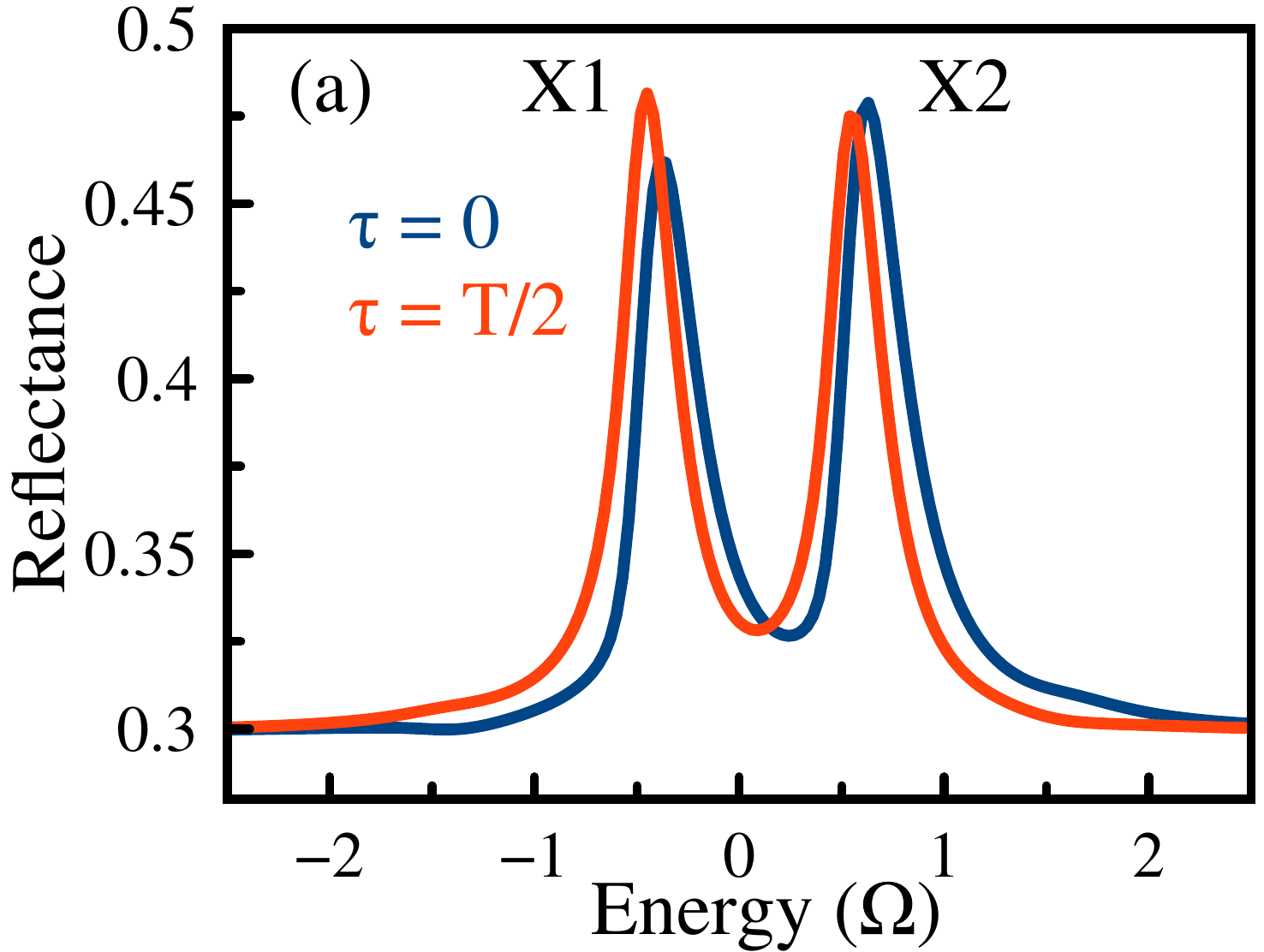}}
 \end{minipage}
 \hfill
  \begin{minipage}[h]{0.49\linewidth}
 \center{\includegraphics[width=\columnwidth]{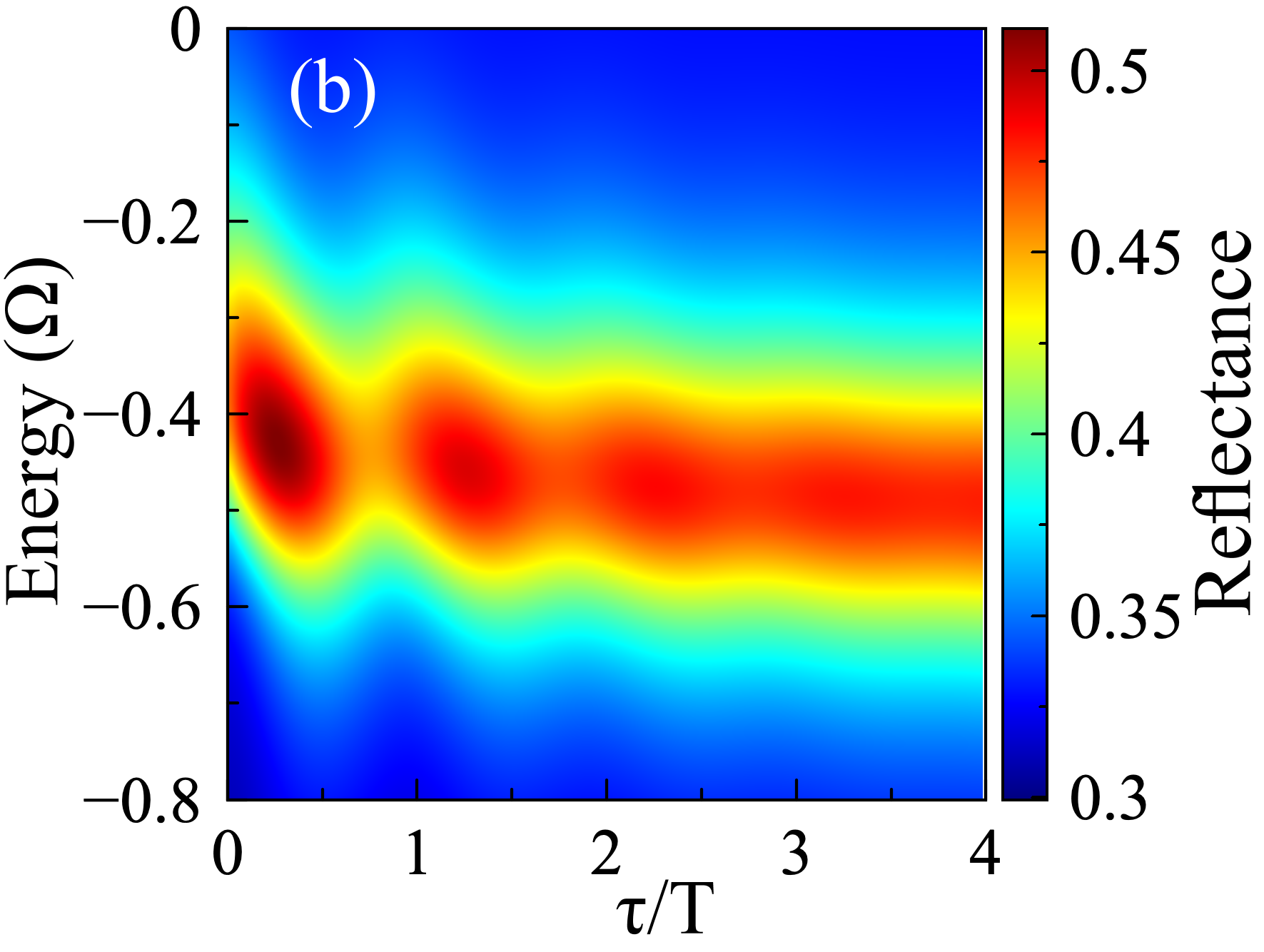}}
 \end{minipage}
 \vfill
   \begin{minipage}[h]{0.49\linewidth}
 \center{\includegraphics[width=\columnwidth]{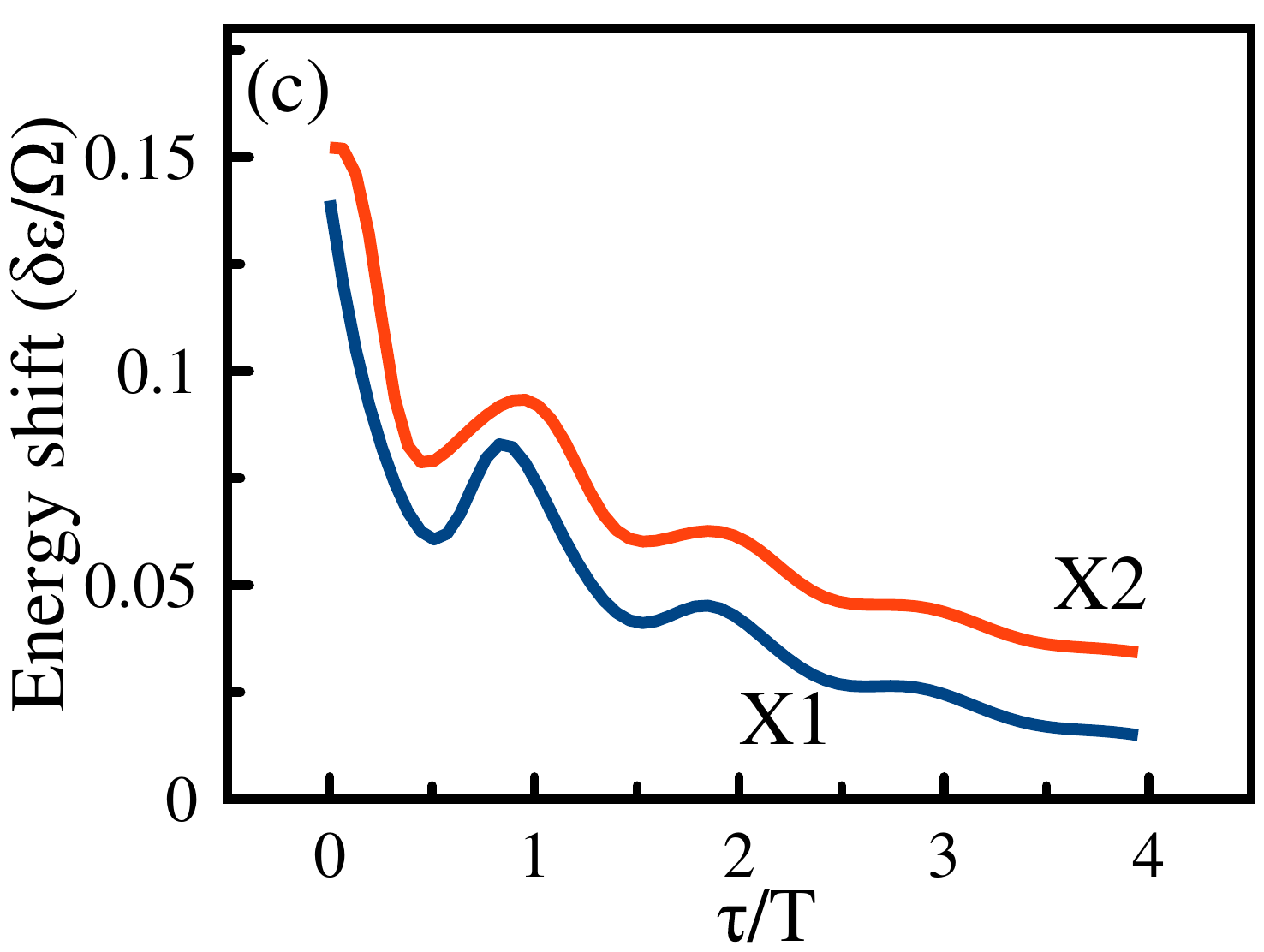}}
 \end{minipage}
 \hfill
   \begin{minipage}[h]{0.49\linewidth}
 \center{\includegraphics[width=\columnwidth]{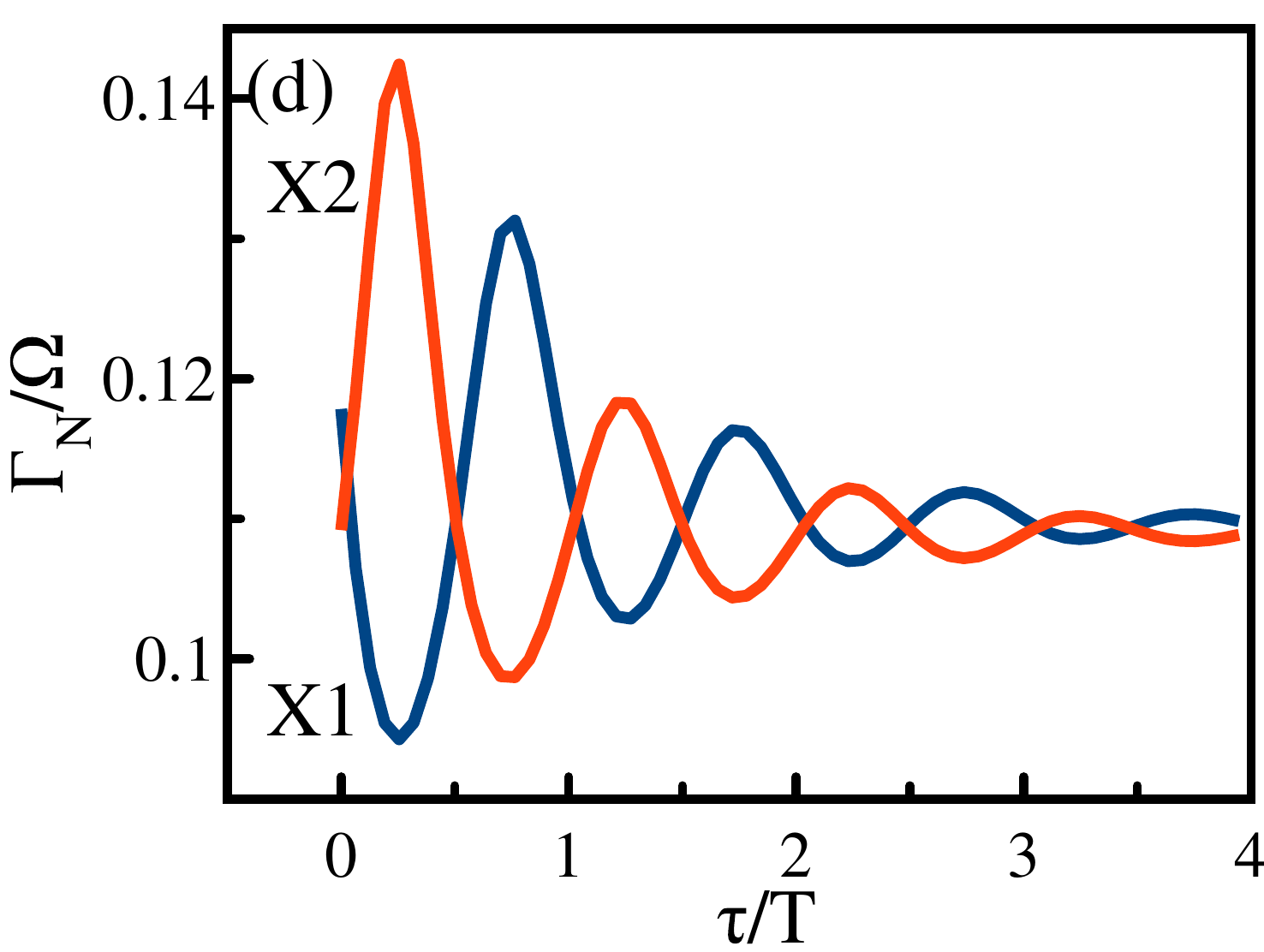}}
 \end{minipage}
 \caption{
(a) The calculated reflectance spectra for the delays $\tau=0$ and $\tau=T/2$ (blue and red curves respectively). (b) Color plot of calculated delay dependence of reflectance spectrum for the parameters $\gamma_0=\gamma_c=0.16\Omega$, $\omega_{ex} N_0 = 0.16\Omega$, $r_{01}=-0.54$. (c) Delay dependence of the energy shift of X1 (red) and X2 (blue) resonances determined from the fit of reflectance in (b) (red curve is shifted by 0.02 for clarity). (d) Delay dependence of non-radiative broadening $\Gamma$ of X1 (red) and X2 (blue) resonances.
} 
 \label{CalcParam}
 \end{figure}

To qualitatively describe the effect of the oscillating exchange interaction strength, we note that, in the real space, the quantum beats described by Eq.~\eqref{eq:beats} represent the oscillations of the pump exciton density between the two sides of the QW, see Figure~\ref{Fig1}(c). The density of the probe excitons oscillates in a similar way, but with the time delay $\tau$. These oscillations of the spatial density of electrically neutral excitons are analogous to the charge jumps in the Hertzian dipole~\cite{HertzDip1, HertzDip2}. If the phase of these oscillations is the same, i.e. $\tau=mT$ ($m=0,1,2,\ldots$) with $T=2\pi/\Omega$, than the maxima of pump and probe excitons always overlap, and their interaction leads to the effective increase of the exciton resonance energy. In the opposite case of $\tau=mT+T/2$, the pump and probe excitons oscillate between the sides of the QW with the opposite phases and hardly overlap, see Figure~\ref{Fig1}(c). In this case the probe exciton resonance energies $\omega_{1,2}$ remain ``bare''. The shift of the resonance frequencies is demonstrated clearly in Figure~\ref{CalcParam}(a), where the reflectance spectra for the delays $\tau=0$ and $T/2$ are shown.

To describe the full dependence of the exciton resonance energies on $\tau$, we calculate the spectra for different delays~\cite{Suppl} and fit them in the same manner as the experimental ones. The result of this calculation is shown in Figure~\ref{CalcParam}(b). In panels (c) and (d) one can see the in-phase delay oscillations of the X1 and X2 exciton resonance energy shifts and anti-phase oscillations of the non-radiative broadening exactly as found in experiments (compare Figure~\ref{FitParam} and Figure~\ref{CalcParam}). From panel (c) one can see, that the energy shift oscillates around the value determined by the incoherent exciton-exciton interaction. The decay of the pump exciton population leads to the decay of the amplitude of energy oscillations as well as well as to the decay of the amplitude of the non-oscillating shift. 


In conclusion, we observed coherent nonlinear exciton dynamics in a wide InGaAs QW. We modified the pump-probe method to include the spectral resolution and measured the modification of the parameters of the exciton resonances as functions of the delay between the pump and probe pulses. We found  delay-dependent oscillations of the exciton resonance energies. These oscillations arise from the real-space exciton density oscillations caused by quantum beats in a manner similar to a Hertzian dipole. The oscillating density results in a delay dependent interaction strength between pump and probe excitons excited into the superposition state. Our results opens up the way  to investigate coherent quantum superposition states through the incoherent interaction in any quantum multilevel system with strong interaction between excitations.

The authors are grateful to M. M. Glazov, M. V. Durnev and I. V. Ignatiev for fruitful discussions. The authors acknowledge SPbU for research grant 11.34.2.2012 (ID 28874264) and the Russian-German collaboration in the frame of CRC TRR 160 project A8 (grant number 249492093) supported by DFG and RFBR grants 19-52-12032 and 19-52-12038. A.V.T. and A.S.K. acknowledge RFBR grants 20-32-70131 and 18-32-00516.  D.S.S. was partially supported by the RF President Grant No. MK-1576.2019.2 and the Basis Foundation. The theoretical modelling was supported by the Russian Science Foundation (Grant No. 19-72-00081). I. Ya. G. acknowledges RFBR grant 19-02-00576a.
The authors thank Recourse Center "Nanophotonics" for providing the heterostructure studied in present work.

\end{document}